# Asymmetric Andreev resonant state with a magnetic exchange field in spin-triplet superconducting monolayer $MoS_2$

H. Goudarzi[1], M. Khezerlou[1,2], and S.F. Ebadzadeh[1]

[1]*Department of Physics, Faculty of Science, Urmia University, Urmia, P.O.Box: 165, Iran*
[2]*National Elites Foundation, Iran*

**Abstract**

Featuring spin-valley degree of freedom by a magnetic exchange field-induction to gain transport of charge carriers through a junction based on superconducting subgap tunneling can provide a new scenario for future electronics. Transmission of low-energy Dirac-like electron (hole) quasiparticles through a ferromagnet/superconductor (F/S) interface can be of noticeable importance due to strong spin-orbit coupling in the valence band of monolayer $MoS_2$ (ML-MDS). The magnetic exchange field (MEF) of a ferromagnetic section on top of ML-MDS may affect the electron (hole) excitations for spin-up and spin-down electrons, differently. Tuning the MEF enables one to control either electrical properties (such as band gap, SOC and etc.) or spin-polarized transport. We study the influence of MEF on the chirality of Andreev resonant state (ARS) appearing at the relating F/S interface, in which the induced pairing order parameter is chiral $p$-wave symmetry. The resulting normal conductance is found to be more sensitive to the magnitude of MEF and doping regime of F region. Unconventional spin-triplet $p$-wave symmetry features the zero-bias conductance, which strongly depends on $p$-doping level of F region in the relating NFS junction.

## 1 INTRODUCTION

Two-dimensional layered transition metal dichalcogenides (TMDC) including Dirac-like charge carries, like graphene, present distinct peculiar physical and dynamical properties[1, 2]. Particularly, in monolayer molybdenum disulfide the charge carriers demonstrate either electron-like or hole-like quasiparticles belonging to two inequivalent degenerate valleys[3, 4]. ML-MDS has a direct band gap and, therefore, is capable for electronics and optical applications. The charge carrier mobility is over $200\ cm^2/V.s$ at room temperature. Specifically, the inversion symmetry breaks in monolayer $MoS_2$, and two inequivalent valleys are interconnected by time reversal symmetry [5, 6]. These features and also strong spin-orbit coupling (SOC) are responsible to the spin-valley degree of freedom in charge transport. The coexistence of valley and spin Hall effect is resulted from coupling of spin and momentum space (valley) at the valence band edges [6]. Consequently, the strong SOC originated from the heavy atom of molybdenum may play an essential role in spin-related investigations. In this regard, presence a magnetic exchange field via the proximity-induction significantly influences the phenomena related to the spin-splitting band structure. This is very useful, since it provides accumulations of spin and valley-polarized carriers with long relaxation times. The valley and spin polarization owing to the MEF-induction may involve valleytronic and quantum computing applications [7, 8, 9].

Furthermore, it is shown that the control of valley polarization can be possible in ML-MDS by removing the valley degeneracy [10, 11, 12]. The applied exchange splitting energy to the ML-MDS results in a large spin splitting in $K'$ valley in the valence band, leading to a novel behavior of pseduo-relativistic

Klein tunneling. In our previous work [13], the effect of an exchange field on Klein tunneling and resulting magnetoresistance was studied in a ferromagnet/insulator/ferromagnet junction. Newly, utilizing a magnetic field is shown to Zeeman split the band edge states in different valleys [14, 15, 16]. In TMDC, utilizing interfacial MEF can overcome the small valley splitting, and it results in breaking time reversal symmetry[7, 8]. Zhao and et al. [17] showed an enhanced valley splitting in monolayer $WSe_2$ utilizing the MEF from a ferromagnetic EuS substrate. Also, they have found that the magneto-reflectance measurement shows at $1\,T$ magnetic field a valley splitting of $2.5\,meV$.

The proximity-induced superconductivity and ferromagnetism have been experimentally shown in ML-MDS [18, 19, 20, 21, 22, 23, 24, 25, 26, 27]. Recently, several investigations have been reported on the ML-MDS superconductor junctions regarding Andreev process at the interface [28, 29, 30, 31, 32]. In this paper, we reveal the formation of Andreev resonant energy and resulting tunneling conductance at the F/S interface, where the superconducting order parameter is taken to be spin-triplet $p$-wave symmetry. However, there is an essential physics in this system. Specifically, in similar situation in topological insulators, the magnetization induction to the surface states leads to open a band gap at Dirac point, and resulting chiral Majorana mode energies appear at the F/S interface [33, 34, 35, 36]. The chirality of Majorana modes is provided by the perpendicular component of induced magnetization. This is considered as an interesting feature of magnetization effect in two-dimensional newly discovered materials. We find a new effect of MEF on the Andreev resonant state (ARS). The chirality symmetry breaking of ARS happens in the presence of MEF. We consider the essential dynamical band parameters of ML-MDS contributions (SOC interaction $\lambda$, electron-hole mass asymmetry term $\alpha$ and topological term $\beta$) to the Andreev process. Moreover, due to valence band spin-splitting in ML-MDS caused by strong SOC, the doping regime is a notable aspect to control the transport of charge carriers (for example, see Refs. [28, 29, 30, 31, 32]). This feature presents more considerable, when spin-splitting is highlighted by a MEF. Indeed, it needs, for experimental applications, to determine the range of permissible doping in F region. This paper is organized as follows. Sec. 2 is devoted to present the proposed model and formalism to obtain the exact form of $MoS_2$ superconducting dispersion energy and corresponding wavefunctions. The normal and Andreev reflection coefficients are found by matching the wavefunctions at the interface. The numerical results of ARS and resulting tunneling conductance considering the strong spin-valley effect caused by ferromagnetic exchange field and also chiral superconducting order parameter are presented, and their main characteristics are discussed in sections 3. Finally, we close with a brief summary.

## 2 THEORETICAL FORMALISM

The low energy band structure of ML-MDS can be described by the modified Dirac Hamiltonian [37]. This Hamiltonian in addition to first order term of momentum for 2D massive fermions, contains the quadratic terms originated from the difference mass between electron and hole $\alpha$, and also topological characteristics $\beta$. The strong spin-orbit coupling leads to distinct spin splitting at the valence band for different valleys. In the presence of an exchange field $h$ and superconducting gap induced by proximity effect, the Dirac-Bogoliubov-de Gennes (DBdG) Hamiltonian is given by:

$$\mathcal{H} = \begin{pmatrix} h_0 - E_F + U(x) - \iota h & \Delta_S(\mathbf{k}) e^{i\varphi} \\ \Delta_S^*(\mathbf{k}) e^{-i\varphi} & -h_0 + E_F - U(x) - \iota h \end{pmatrix}, \quad (1)$$

where $h_0 = \hbar v_F \mathbf{k} \cdot \boldsymbol{\sigma}_\tau + \Delta \sigma_z + \lambda s \tau (1 - \sigma_z) + \frac{\hbar^2 |k|^2}{2m_0}(\frac{\alpha}{2} + \frac{\beta}{2}\sigma_z)$ is the full effective Hamiltonian of monolayer $MoS_2$, and $\boldsymbol{\sigma}_\tau = (\tau \sigma_x, \sigma_y)$ are the Pauli matrices. The spin-up and spin-down is labeled by $s = \pm 1$, and valley index $\tau = \pm 1$ denotes the $K$ and $K'$ valleys. The bare electron mass is $m_0 = 0.05 \times 10^{-10}\,(eVs^2/m^2)$, and topological and mass difference band parameters are evaluated by $\beta = 2.21$ and $\alpha = 0.43$, respectively. $\Delta$ is the direct band gap, $\lambda \approx 0.08\,eV$ and $v_F = 0.53 \times 10^6\,m/s$ denote the spin-orbit coupling and Fermi velocity, respectively. The electrostatic potential $U(x)$ gives the relative shift to the Fermi energy $E_F$ in normal N, ferromagnet F and superconductor S regions as $\mu_{N,F,S} = E_F - U(x)$, which denotes the chemical potential in each region. $\Delta_S(\mathbf{k})$ is the superconducting order parameter. The globally broken $U(1)$ symmetry in the superconductor is characterized with phase $\varphi$. Obviously, the spatial part of triplet order parameter is an odd function under exchange of momentum of the two particles, while the spin part is even. For a spin-triplet symmetry the order parameter is expressed using the $d$-vector as:

$$\Delta_{s,s'}(\mathbf{k}) = [\mathbf{d}(\mathbf{k}) \cdot \tau] i\tau_y \tag{2}$$

where $\mathbf{d}$, $\tau$, and $\tau_y$ are an odd-parity function of $\mathbf{k}$ and Pauli matrices, which describes the real electron spin $(s, s')$, respectively. The direction of $\mathbf{d}$-vector is perpendicular to total spin $S$ of a Cooper pair. This order symmetry has a off-diagonal components. Without loss of generality, let us consider the case of $\mathbf{d}(\mathbf{k}) = \Delta_S(\mathbf{k})\hat{z}$, which means $k \perp S$. The symmetry of ML-MDS lattice plays, of course, a central role in $\mathbf{k}$-dependency of $\mathbf{d}$. In ML-MDS, the promising pairing symmetry is $p_x$ and chiral $p_x + ip_y$-wave symmetries [38]. Diagonalizing Eq. (1) produces the following energy-momentum dispersion around Dirac point [29]:

$$\varepsilon_S = \xi \sqrt{(|\Delta_S|\eta)^2 + \left(\frac{A+B}{2} + v\sqrt{\frac{(A-B)^2}{4} + v_F^2|k_S|^2 + |\Delta_S|^2(1-\eta^2)}\right)^2} \tag{3}$$

where $A = \Delta + \frac{|k_S|^2}{2m_0}(\frac{\alpha}{2} + \frac{\beta}{2}) - E_F + U(x)$, $B = -\Delta + 2\lambda s + \frac{|k_S|^2}{2m_0}(\frac{\alpha}{2} - \frac{\beta}{2}) - E_F + U(x)$ and $\eta = \sqrt{1 - (\frac{A-B}{A+B})^2}$. The index $\xi = \pm 1$ denotes the electron-like and hole-like excitations, while $v = \pm$ distinguishes between the conduction and valence bands. Straightforwardly, it is shown that the superconducting order parameter $\Delta_S(k)$ in Eq. (3) is renormalized by chemical potential $\mu_S$, and also appears as an ordinary gap. This electron-hole superconducting excitation is qualitatively different from that obtained for conventional singlet superconductivity [30], so that it seems to remain semigapless. The mean-field conditions are satisfied as long as $\Delta_S \ll \mu_S$. In this condition, the exact form of superconductor wavevector of quasiparticles can be acquired from the relation

$$k_S = \frac{1}{v_F \hbar}(k_0 + ik_1), \quad k_0 = \sqrt{AB}$$

where $k_1$ can be responsible to exponentially decaying. In particular, we retain the contribution of $\alpha$ and $\beta$ terms representing one of the essential physics of monolayer $MoS_2$. Hamiltonian (1) can be solved to obtain the wave function for superconductor region. The wavefunction, which includes a contribution from both electron-like and hole-like quasiparticles are found as

$$\psi_S^e = \begin{pmatrix} \zeta\beta_1 \\ \zeta\beta_1 e^{i\tau\theta_S} \\ e^{-i\gamma^e} e^{-i\varphi} e^{i\tau\theta_S} \\ e^{-i\gamma^e} e^{-i\varphi} \end{pmatrix} e^{i\tau k_S^x x}, \quad \psi_S^h = \begin{pmatrix} \zeta\beta_2 \\ -\zeta\beta_2 e^{-i\tau\theta_S} \\ -e^{-i\gamma^h} e^{-i\varphi} e^{-i\tau\theta_S} \\ e^{-i\gamma^h} e^{-i\varphi} \end{pmatrix} e^{-i\tau k_S^x x}, \tag{4}$$

where we introduce $\beta_{1(2)} = -\frac{\varepsilon}{\Delta_S} - (+)\sqrt{\frac{\varepsilon^2}{|\Delta_S|^2} - \eta^2}$, $\zeta = \frac{A+B}{2\sqrt{AB}}$, $e^{i\gamma^{e,h}} = \frac{\Delta_S(\mathbf{k})}{|\Delta_S(\mathbf{k})|}$. Fermi level of each region can be tuned by the magnitude of chemical potential. The momentum of ferromagnetic electrons is coupled with the exchange field in $F$ region. Therefore, the corresponding Fermi wavevector needs to be acquired from its eigenstate:

$$k_{FF} = \pm \sqrt{\frac{\omega_1 \frac{\alpha}{2} + \omega_2 \frac{\beta}{2} + m_0 v_F^2 - \sqrt{(\omega_1 \frac{\beta}{2} + \omega_2 \frac{\alpha}{2})^2 + (m_0 v_F^2)^2 + 2m_0 v_F^2(\omega_1 \frac{\alpha}{2} + \omega_2 \frac{\beta}{2})}}{\frac{\hbar^2}{2m_0}(\frac{\alpha}{2}^2 - \frac{\beta}{2}^2)}}, \tag{5}$$

where we define $\omega_1 = -E_{FF} - \lambda s\tau + \iota h$, $\omega_2 = -\lambda s\tau + \Delta$. The wave functions in ferromagnetic region are given by:

$$\psi_F^{e+} = \begin{pmatrix} 1, \tau e^{i\tau\theta_F^e} A_F^e, 0, 0 \end{pmatrix}^T e^{i\tau k_F^{ex} x}, \quad \psi_F^{h-} = \begin{pmatrix} 0, 0, 1, -\tau e^{i\tau\theta_F^h} A_F^h \end{pmatrix}^T e^{-i\tau k_F^{hx} x},$$

where

$$A_F^{e(h)} = \hbar v_F \left|k_F^{e(h)}\right| / \left((-)\epsilon - E_{FF} + \Delta - 2\lambda s\tau - \frac{\hbar^2 \left|k_F^{e(h)}\right|^2}{2m_0}(\frac{\alpha}{2} - \frac{\beta}{2}) + \iota h\right).$$

Having established the states that participate in the scattering, the total wave function for a right-moving electron with angle of incidence $\theta^e$, a left-moving electron by the substitution $\theta^e \longrightarrow \pi - \theta^e$ and a left-moving hole by angle of reflection $\theta^h$ may then be written as:

$$\psi_N = \frac{1}{\sqrt{\mathcal{N}^e}} \begin{pmatrix} 1 \\ \tau e^{i\tau\theta_N^e} A_N^e \\ 0 \\ 0 \end{pmatrix} e^{i\tau k_N^{ex} x} + \frac{R(\epsilon, \theta^e)}{\sqrt{\mathcal{N}^e}} \begin{pmatrix} 1 \\ -\tau e^{-i\tau\theta_N^e} A_N^e \\ 0 \\ 0 \end{pmatrix} e^{-i\tau k_N^{ex} x} +$$

$$+ \frac{R_A(\epsilon, \theta^e)}{\sqrt{\mathcal{N}^h}} \begin{pmatrix} 0 \\ 0 \\ 1 \\ \tau e^{-i\tau\theta_N^h} A_N^h \end{pmatrix} e^{i\tau k_N^{hx} x}, \quad (6)$$

where we define $A_N^{e(h)} = \hbar v_F \left|k_N^{e(h)}\right| / \left((-)\epsilon - E_{FN} + \Delta - 2\lambda s\tau - \frac{\hbar^2 \left|k_N^{e(h)}\right|^2}{2m_0}(\frac{\alpha}{2} - \frac{\beta}{2})\right)$. $R(\epsilon, \theta^e)$ and $R_A(\epsilon, \theta^e)$ are amplitude of normal and Andreev reflections, respectively. The normalization factor $\mathcal{N}^{e(h)}$ ensure that the particle current density is conserved, which has been given in previous work [13].

It is instructive to consider the effect of Fermi vector mismatch on the Andreev reflection amplitude. To explore how the Fermi wave vector mismatch influences the scattering processes, the Fermi momentum in normal and superconducting regions of the system may be controlled by means of tuning the chemical potential. We proceed to study the Andreev reflection and resulting conductance in F/S and NFS ML-MDS junctions. The normal and ferromagnetic regions are extended from $x < 0$ and $x = 0$ to $x = d$, respectively. ML-MDS is covered by superconducting electrodes in the region $x = d$ to $x = \infty$. The strategy for calculating the scattering coefficients in the junction is to match the wave functions at boundaries $\psi_N|_{x=0} = \psi_F|_{x=0}, \psi_F|_{x=d} = \psi_S|_{x=d}$, where $\psi_S = t^e \psi_S^e + t^h \psi_S^h$. The coefficients $t^e$ and $t^h$ correspond to transmission of electron and hole, receptively. We find the following solution for the normal and Andreev reflection coefficients:

$$R_A(\epsilon, \theta^e) = \left(\frac{\mathcal{N}^e}{\mathcal{N}^h}\right) \mathcal{M} \left(e^{-i\gamma^h} e^{-i\gamma^e} \Gamma_6 \Gamma_{11} + e^{-i\gamma^h} e^{-i\gamma^e} \Gamma_5 \Gamma_{12}\right),$$

$$R(\epsilon, \theta^e) = \mathcal{M} \left(\xi\beta_1 e^{-i\gamma^h} \Gamma_6 \Gamma_9 + \xi\beta_2 e^{-i\gamma^e} \Gamma_5 \Gamma_{10}\right) - 1,$$

The parameters of $\Gamma$ are introduced in Appendix. The above expressions demonstrate exactly the Andreev process at the interface, leading to the spin-valley polarized transport of charge carriers through relevant junctions.

## 3 NUMERICAL RESULTS AND DISCUSSION

In this section, we analyze in detail the dynamical transport properties of FS and NFS junctions when the $p$-wave symmetry pairing is deposited on top of ML-MDS. Coupling the magnetic exchange field in F region with strong SOC of ML-MDS influences the Andreev resonant states at the F/S interface. On the other hand, the unconventional superconducting order with nonzero orbital angular momentum ($l = 1$) plays a crucial role in materials with strong SOC. Recently, it is shown that the quasiparticle superconducting excitations are influenced by the triplet component of $p$-wave pairing symmetry [29].

Also, an exchange field induction to ML-MDS is found to give rise a notable spin-splitting energy in valence band between spin-up and spin-down charge carriers, leading to the spin-valley polarized Klein tunneling [13]. In this regards, if we consider a F/S interface, the above features can straightforwardly determine the exhibition of Andreev process and resulting subgap tunneling conductance. A typical F/S interface can be of significant importance owing to the appearance of Andreev resonant states and, especially, formation of chiral Majorana mode in topological insulators [33, 34, 35, 36].

## 3.1 Andreev reflection and resonant state

In one-dimensional limit, transport of tunneling electrons is in $x$-direction with wavevector $k^x$ and incident angle $0 \leq \theta_N^e \leq \pi/2$. Conservation of momentum in $y$-direction enables one to find incident electron (hole) angle from S region to the interface $\theta_S^{e(h)}$ in terms of $\theta_N^e$, which is given by $\theta_S^{e(h)} = \arcsin(k/k_S \sin \theta_N^e)$. We show in Fig. 1 the angle-resolved Andreev and normal reflection probabilities in zero-bias $\epsilon(eV) = 0$, where the MEF affects the Andreev process. Similar to previously obtained results [28, 29, 30, 32], the band parameters $\lambda$ and $\beta$ of ML-MDS make a significant effect on the reflection of quasiparticles from F/S interface. In the presence of MEF, there is no perfect AR at normal incidence. However, the density of probability of reflections is conserved, i.e., $|R(\epsilon, \theta^e)|^2 + |R_A(\epsilon, \theta^e)|^2 = 1$. The signature of exchange field is believed to decline AR in NFS junction, since the incident electrons from N region experience a high spin-splitting energy depending on their spin-polarization. In this system, the effective superconductor subgap is adjusted by a factor $\eta$ including the characteristic band parameters of ML-MDS.

Next, we proceed to look for the possibility to form chiral bound energy mode resulted from Andreev resonant state at the F/S interface. Occurring the perfect AR, the electron reflected to N region may vanish. Thus, the ARS reads:

$$\tilde{\varepsilon}(\theta_N^e, h) = \frac{\eta |\Delta_S|}{\sqrt{c^2+1}}; \quad c = \tan\left[\frac{1}{2i}\ln\left(\frac{-\tau_1}{\tau_2}\right)\right],$$

where $\tau_1 = \Gamma_5 e^{-i\gamma^e}(2A_N^e \cos\theta_N^e \Gamma_{10} - \Gamma_8)$, $\tau_2 = \Gamma_6 e^{-i\gamma^h}(2A_N^e \cos\theta_N^e \Gamma_9 - \Gamma_7)$. Parameters $\Gamma$ are given in Appendix. In Fig. 2, we present these states in terms of incident electron from N region for several values of MEF. As a remarkable point, we find the chirality symmetry of ARS to be conserved in the absence of MEF, whereas it is broken in the presence of MEF, so that, the magnitude of ARS differs in its positive and negative values (left and right sides of curves $\Delta\tilde{\varepsilon}(\theta_N^e, h)$) for nonzero incident angles. Comparing to the gapless surface state of topological insulator, where chiral Majorana mode energy appears at the F/S interface[33, 34, 35, 36], the direct band gap of ML-MDS can be responsible to disappearing the chirality symmetry of ARS. However, in topological insulator, the perpendicular component of magnetization ($m_z$) causes to open a gap in Dirac point, and the sign of magnetization provides chiral Majorana mode of Andreev bound state.

## 3.2 Subgap normal conductance

We now proceed to calculate the subgap tunneling conductance in NFS and FS junctions. According to the Blonder-Tinkham-Klapwijk formalism [39], we can calculate the tunneling conductance by

$$G(eV) = \sum_{s,\tau=\pm 1} G_0^{s,\tau} \int_0^{\theta_c} \left(1 - |R(\epsilon,\theta^e)|^2 + |R_A(\epsilon,\theta^e)|^2\right) \cos\theta^e d\theta^e, \tag{7}$$

where $G_0^{s,\tau} = e^2 N_{s,\tau}(eV)/h$ is the ballistic conductance of spin and valley-dependent transverse modes $N_{s,\tau} = kw/\pi$ in a sheet of $MoS_2$ of width $w$ that $eV$ denotes the bias voltage. The upper limit of integration in Eq. (7) needs to obtain exactly based on the fact that the incidence angle of electron-hole in the three regions may be less than $\frac{\pi}{2}$.

Dependence of the resulting normal Andreev conductance $G/G_0$ on MEF is presented in Figs. 3 for three different values of MEF $h = 0.0\lambda, 0.5\lambda, 1.0\lambda$ for $p$-doped F region. For both FS and NFS junctions the appearance of zero-bias conductance is found. This is the main feature of unconventional superconducting order parameter [40]. We observe that ,at a determined bias energy, the conductance curve asymptotically vanishes and presents a peak. This can be described by the fact that the effective superconducting gap in electron-hole energy excitations is renormalized by a coefficient $\eta$. In NFS junction, the valence band spin-splitting difference between N and F regions gives rise to momentum difference of Andreev reflected electron (hole), and thus, it leads to diminish AR probability and resulting subgap conductance, as seen in Fig. 3(b). Whereas, in FS junction, the conductance peak grows up with the increase of MEF (see, Fig. 3(a)), because the SOC is dominated, here. This is in contrast to the similar situation in graphene junction when $h < \mu_F$ [41]. The above result can be understood by the fact that, in one hand, the effective superconducting gap with spin triplet-wave symmetry is enhanced by the SOC interaction and, on the other hand, coupling MEF with momentum of Andreev reflected electron (hole) leads to decrease the wavevector mismatch.

Fig. 4(b) shows that the zero-bias conductance (ZBC) strongly depends on $p$-doping level of F region in NFS junction, while it is almost constant $G/G_0 \cong 0.73$ in FS junction, as seen in Fig. 4(a). For a constant exchange field $h = 0.5\lambda$, the conductance peak enhances with the increase of chemical potential of F region $\mu_F$ for $\epsilon\,(eV) \cong 0.1\,\Delta_S$, in NFS structure. In FS structure, we observe a contrary behavior herein. For $p$-doped F region ($\mu_F = -0.95\,eV$), the valence band spin-splitting is strongly influenced by the MEF, and there are four possible critical values for band energy ($-\Delta/2 \pm 2\lambda \pm h$), which allows appropriately to occur AR in the range of ferromagnetic chemical potential $-\Delta/2 - 2\lambda - h < \mu_F < -\Delta/2 + 2\lambda + h$. Indeed, this antithetical outcomes are directly resulted from Andreev process at the N/S or F/S interfaces. To more clarify, one can investigate Andreev bound stats in a SFS Josephson junction as a future work. In FS junction, the switching behavior of normal conductance can be controlled by tuning the bias energy. In low $p$-doping of F region, the switching is shifted toward zero bias. With the increase of bias energy up to effective superconducting gap $\eta\Delta_S$, the tunneling conductance results in a constant $\sim 0.76$ in FS and $\sim 0.65$ in NFS junctions. In the latter case, the doping of F region has no effect on the conductance. Finally, we focus on the influence of dynamical band parameters of ML-MDS such as electron-hole mass asymmetry term $\alpha$, topological term $\beta$ and SOC term $\lambda$ in the resulting subgap conductance in NFS junction, as seen in Fig. 5. In the case of MEF = $0.5\lambda$ and $p$-doped N and F regions ($\mu_F = -0.97\,eV, \mu_N = -0.95\,eV$), the term $\alpha$ has no effect, as in agreement with previous works [28, 29, 30, 32]. The absence of topological term $\beta$ gives rise to enhancing the conductance peak. More importantly, taking the SOC term $\lambda$ to neglect, we find a sharp switching conductance in zero bias.

## 4 CONCLUSION

In summary, we have considered transition metal dichalcogenide $MX_2$ ($M = Mo, X = S$) in layered structure, where Dirac-like electrons (holes) have experienced proximity-induced a magnetic exchange field or a superconducting pair potential. The unconventional spin-triplet $p$-wave order parameter has been found to be more effective in our proposed structure owing to the existence of a strong SOC in band structure of monolayer $MoS_2$. Spin-valley degree of freedom is a key point to control the transport of charge carriers by tuning the valence band locked-spin-valley splitting via the applying a MEF. A key finding of the present work is that the Andreev resonant energy at the relating F/S interface exhibits an asymmetric behavior with the presence of MEF. Thus, we can not consider the chirality symmetry of ARS, comparing to the similar situation in topological insulator F/S interface with inducing a magnetization perpendicular to the surface. Andreev process at the F/S interface has led to the tunneling conductance, which was controlled by tuning the MEF and also doping regime. The wavevector mismatch has been studied to obtain the relating normal conductance in FS and NFS junctions. A sharp switching conductance in zero bias has been achieved in the absence of SOC.

**Acknowledgment**

This work was supported by National Elites Foundation of I.R.Iran. Authors acknowledge the Vice-Presidency for Science and Technology of I.R.Iran.

## APPENDIX

The parameters $\Gamma$ in Andreev and normal reflections are given as:

$$\mathcal{M} = \frac{2\tau A_N^e \cos(\tau\theta_N^e)}{\xi\beta_1 e^{-i\gamma^h}\Gamma_6\Gamma_7 + \xi\beta_2 e^{-i\gamma^e}\Gamma_8\Gamma_5},$$

$$\Gamma_1 = \frac{A_F^e e^{i\tau\theta_F^e} - e^{i\tau\theta_S}}{2A_F^e \cos(\tau\theta_F^e)}, \quad \Gamma_2 = \frac{A_F^e e^{i\tau\theta_F^e} + e^{-i\tau\theta_S}}{2A_F^e \cos(\tau\theta_F^e)},$$

$$\Gamma_3 = \frac{1 + e^{i\tau\theta_S} e^{i\tau\theta_F^h} A_F^h}{2A_F^h \cos(\tau\theta_F^h)}, \quad \Gamma_4 = \frac{1 - e^{-i\tau\theta_S} e^{i\tau\theta_F^h} A_F^h}{2A_F^h \cos(\tau\theta_F^h)},$$

$$\Gamma_5 = \left(A_N^h e^{-i\tau\theta_N^h} + A_F^h e^{i\tau\theta_F^h}\right)\left(e^{i\tau\theta_S} - \Gamma_3\right)e^{i\tau k_F^{hx} d} + \left(A_N^h e^{-i\tau\theta_N^h} - A_F^h e^{-i\tau\theta_F^h}\right)\Gamma_3 e^{-i\tau k_F^{hx} d},$$

$$\Gamma_6 = \left(A_N^h e^{-i\tau\theta_N^h} + A_F^h e^{i\tau\theta_F^h}\right)\left(e^{-i\tau\theta_S} + \Gamma_4\right)e^{i\tau k_F^{hx} d} - \left(A_N^h e^{-i\tau\theta_N^h} - A_F^h e^{-i\tau\theta_F^h}\right)\Gamma_4 e^{-i\tau k_F^{hx} d},$$

$$\Gamma_7 = \left(A_F^h e^{i\tau\theta_F^e} + A_N^e e^{-i\tau\theta_N^e}\right)(1 - \Gamma_1)e^{-i\tau k_F^{ex} d} - \left(A_F^e e^{-i\tau\theta_F^e} - A_N^e e^{-i\tau\theta_N^e}\right)\Gamma_1 e^{i\tau k_F^{ex} d},$$

$$\Gamma_8 = \left(A_F^e e^{i\tau\theta_F^e} + A_N^e e^{-i\tau\theta_N^e}\right)(1 - \Gamma_2)e^{-i\tau k_F^{ex} d} - \left(A_F^e e^{-i\tau\theta_F^e} - A_N^e e^{-i\tau\theta_N^e}\right)\Gamma_2 e^{i\tau k_F^{ex} d},$$

$$\Gamma_9 = (1 - \Gamma_1)e^{-i\tau k_F^{ex} d} + \Gamma_1 e^{i\tau k_F^{ex} d}, \quad \Gamma_{10} = (1 - \Gamma_2)e^{-i\tau k_F^{ex} d} + \Gamma_2 e^{i\tau k_F^{ex} d},$$

$$\Gamma_{11} = \Gamma_3 e^{-i\tau k_F^{hx} d} + (e^{i\tau\theta_S} - \Gamma_3)e^{i\tau k_F^{hx} d}, \quad \Gamma_{12} = \Gamma_4 e^{-i\tau k_F^{hx} d} - (e^{-i\tau\theta_S} + \Gamma_4)e^{i\tau k_F^{hx} d}.$$

**Figure captions**

**Figure 1** (Color online) The probability of normal (dashed lines) and Andreev (solid lines) reflections as a function of incident angle for several values of the magnetic exchange field in NFS junction when $\mu_F = \mu_N = -0.95\ eV$ and $\mu_S = 2\ eV$. It is seen that the maximum Andreev reflection (green solid line) occurs in $h = 0.0\ \lambda$.

**Figure 2** (Color online) The plot shows Andreev resonant state as a function of electron incident angle in NFS junction for several values of MEF when $\mu_N = \mu_F = -1\ eV$ and $\mu_S = 2\ eV$. The solid lines correspond to $h = 0.0\ \lambda$ and the dashed lines $h = 1.0\lambda$ and $1.5\lambda$.

**Figure 3** (Color online) Plot of the tunneling conductance as a function of the bias voltage for several values $h = 0.0\ \lambda$, $h = 0.5\ \lambda$ and $h = 1.0\ \lambda$ in (a) FS junction and (b) NFS junction. we set $\mu_F = -0.95\ eV$ and $\mu_S = 3\ eV$ in Fig.(a) and $\mu_F = -0.98\ eV, \mu_N = -0.95\ eV$ and $\mu_S = 2\ eV$ in Fig.(b).

**Figure 4** (Color online) Normalized Andreev conductance of ML-MDS as a function of bias energy for different values of ferromagnetic chemical potential. (a) Dependence of tunneling conductance for three diferent values $\mu_F = -0.98\ eV$ (green solid line), $\mu_F = -0.95\ eV$ (blue dashed line) and $\mu_F = -0.92\ eV$ (red dashed line) when $h = 0.5\ \lambda$ and $\mu_S = 3\ eV$ in FS junction. (b) Dependence of tunneling conductance for three diferent values $\mu_F = -0.9\ eV$ (green solid line), $\mu_F = -0.95\ eV$ (blue dashed line) and $\mu_F = -0.99\ eV$ (red dashed line) when $h = 0.5\ \lambda$, $\mu_N = -0.95\ eV$ and $\mu_S = 2\ eV$ in NFS junction.

**Figure 5** (Color online) The Plot shows the Andreev conductance as a function of bias voltage in NFS junction. It shows the role of dynamical band parameters of ML-MDS ($\alpha, \beta,$ and $\lambda$). We set $\mu_N = -0.95\ eV, \mu_F = -0.97\ eV, \mu_S = 2\ eV$ and $h = 0.5\ \lambda$.

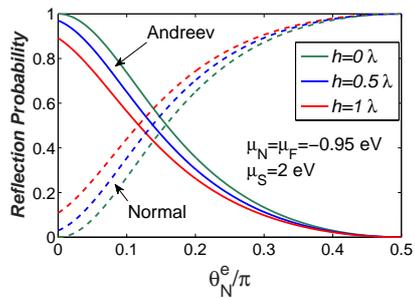

Figure 1:

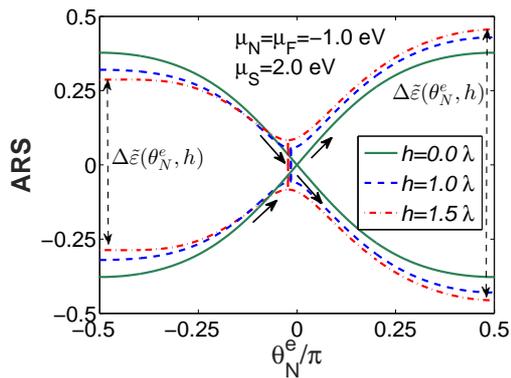

Figure 2:

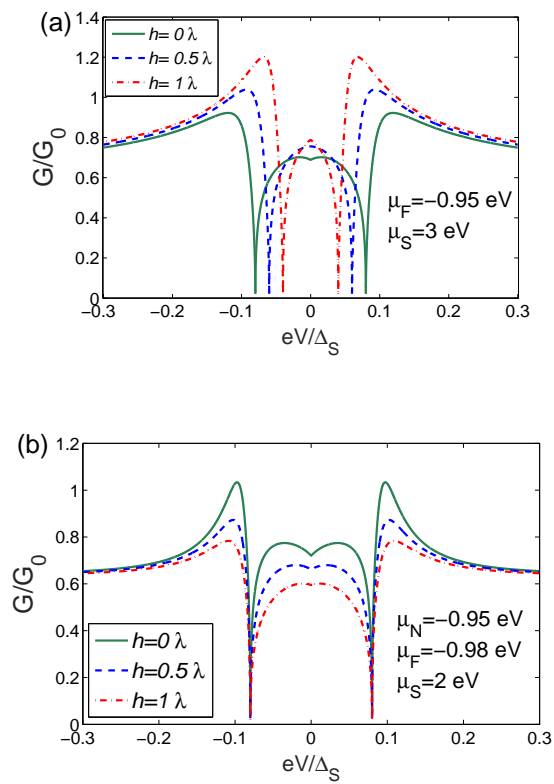

Figure 3:

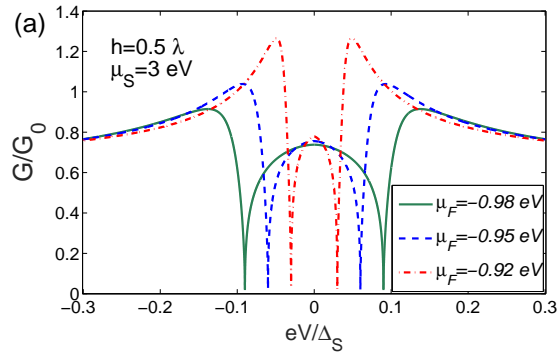

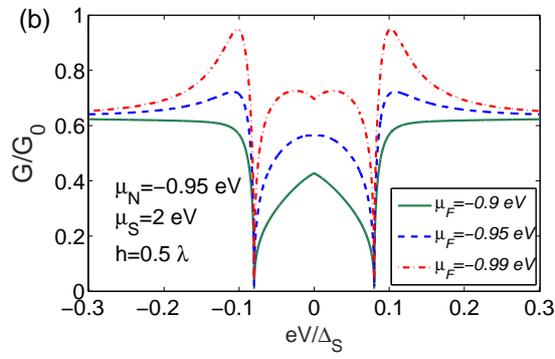

Figure 4:

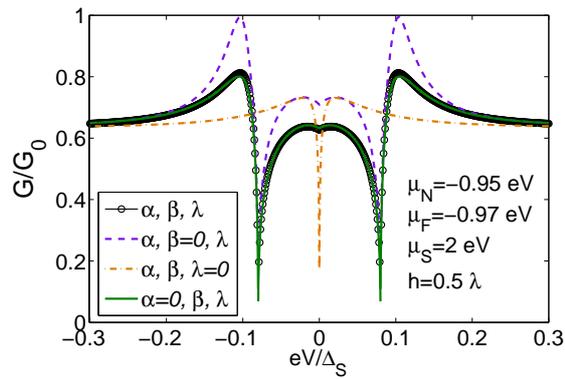

Figure 5: